\documentclass[10pt,a4paper]{article}
\usepackage[utf8]{inputenc}
\usepackage{cite}
\usepackage{hyperref} 
\usepackage{pdfpages}
\usepackage{amsmath}
\usepackage{amsfonts}
\usepackage{amssymb}
\usepackage{float}
\usepackage{bbold}
\usepackage[title]{appendix}
\usepackage{url}
\usepackage{verbatim} 
\usepackage{graphicx}
\usepackage{vmargin}
\DeclareMathAlphabet\mathbfcal{OMS}{cmsy}{b}{n}
\author{J. Antonio Garc\'ia$^1$ and R. Abraham S\'anchez-Isidro$^2$\\
	\small Departamento de F\'isica de Altas Energ\'ias, Instituto de Ciencias Nucleares\\
	\small Universidad Nacional Aut\'onoma de M\'exico,\\
	\small Apartado Postal 70-543, Ciudad de M\'exico, 04510, M\'exico\\
	\small $^1$ garcia@nucleares.unam.mx, $^2$ abraham.sanchez@correo.nucleares.unam.mx}
\title{$\sqrt{T\overline{T}}$-deformed oscillator inspired by ModMax}
\begin{document}
	\maketitle
	\begin{abstract}
	Inspired by a recently proposed Duality and Conformal invariant modification of Maxwell theory (ModMax), we construct a one-parameter family of two-dimensional dynamical system in classical mechanics that share many features with the ModMax theory. It consists of a couple of $\sqrt{T\overline{T}}$-deformed oscillators that nevertheless preserves duality  $(q \rightarrow p,p \rightarrow -q)$ and depends on a continuous parameter $\gamma$, as in the ModMax case. Despite its non-linear features, the system is integrable. Remarkably can be interpreted as a pair of two coupled oscillators whose frequencies depend on some basic invariants that correspond to the duality symmetry and rotational symmetry. Based on the properties of the model, we can construct a non-linear map dependent on $\gamma$ that maps the oscillator in 2D to the nonlinear one, but with parameter $2\gamma$.  
		The dynamics also shows the phenomenon of energy transfer and we calculate a Hannay angle associated to geometric phases and holonomies.
		
	\end{abstract}
	\section{Introduction}
	
Consistent deformation of field theories is a powerful tool to explore in what sense or to what extent a given theory can be deformed by deforming its gauge symmetries and/or couplings in such a way the resulting theory is also consistent (removes the gauge degree of freedom in a consistent way). It is very useful to explore questions about if the theory is unique and/or how it can be extended to new field theories that could be interesting in physics. The deformations are regulated by symmetries and other constraints that we impose according to what questions we are addressing when deforming. All deformations preserve the field content of the given theory. A well known example is a deformation of N free Maxwell fields where the gauge symmetry and the interaction, preserving global Poincaré symmetry, can be deformed to construct the Yang-Mills SU(N) theory. This deformation is also unique under the restrictions imposed 
	\cite{Barnich-Henneaux}.

	We can also deform a free action in many other ways by imposing different restrictions. Maxwell's theory can be deformed in different ways by preserving duality and conformal symmetries in 4D. Some celebrated examples are the Born-Infeld \cite{BI}, Plebansky \cite{plebanski}, and Biankii-Birula deformations \cite{BB-1,BB-2}. Recently a new deformation of Maxwell theory preserving conformal and duality invariance has been constructed \cite{modmax-towsend,Kosyakov}\footnote{The question about the more general Hamiltonian preserving duality and Lorentz invariance was also addressed in \cite{DS}.}. The result is quite interesting. It reduces to previously known deformations and gives us a new non-linear electromagnetic theory (ModMax). Interestingly the theory preserves duality symmetry. As is well known duality invariance is difficult to implement in interacting theories even if we know that the corresponding free theory is duality invariant. Another example of a theory with a duality invariance is the Fierz-Pauli action\cite{T-H}.  Still we do not know how to implement this duality in the full General Relativity. The duality symmetry plays a very important role in the formulation of string theory and many investigations have been developed to understand its implications. Understanding this fascinating symmetry is the focus of many studies in string theories.
	
	The ModMax theory is a one-parameter family of non-linear electrodynamics theories that was constructed in \cite{modmax-towsend}. The Lagrangian derivation was also presented in \cite{Kosyakov}.
	The Legendre transformation between the Lagrangian and Hamiltonian theories is non trivial.
	 ModMax is constructed from the two basic Lorentz invariants of Maxwell theory $S=-(1/4)F^{\mu\nu}F_{\mu\nu}$ and $P=-(1/4)F_{\mu\nu}\tilde{F}^{\mu\nu}$, where $\tilde{F}^{\mu\nu}$ is the Hodge dual of $F_{\mu\nu}$ the usual Maxwell strength field. The conformal symmetry implies that the trace of the energy-momentum tensor is zero and this in turn implies that the Lagrangian is a homogeneous function of degree one in $S$ and $P$. ModMax theory is the unique conformal $U(1)$ gauge theory with a non-linear constitutive relations, which preserves electric/magnetic duality  invariance. In the phase space this means that ModMax is invariant under the $S$-duality transformation $(\mathbf{D}\rightarrow -\mathbf{B}, \mathbf{B}\rightarrow \mathbf{D})$. It is important to stress that, the duality symmetry is only manifest in Hamiltonian formalism. In Lagrangian formalism this invariance is not manifest \cite{Desser}. The standard definition of $\mathbf{D}$ is $\mathbf{D}=\frac{\partial L}{\partial \mathbf{E}}$ and in ModMax the vector $\mathbf{D}$ depends on $\mathbf{E}$ and $\mathbf{B}$ as in a generic non-linear electrodynamics theory \cite{gibbons}. We review some of the basic properties of ModMax theory in appendices \ref{Hamiltonian-ModMax}, and \ref{Lagrangian-ModMax}.
	
	Although, there may be other Lorentz and duality invariant theories of electrodynamics corresponding to other deformations, there exist just two theories that are conformal invariant, Bialynicki-Birula electrodynamics and ModMax \cite{modmax-towsend}.  
	ModMax theory has been scrutinized and extended in recent works \cite{BH-thermodynam-MM,Hamitonian-Roman,Kruglov,Kruglov-MBH,MM-SUSY,MM-charged-particle,Roman-1,Roman-dyons,current-sqrt-flow-MM,nastase,MM-without-linearterm,non-relativistic-MM}.

A very interesting question is if ModMax theory can be constructed from a deformation of Maxwell theory. Fortunately the answer to this question is affirmative and the deformation is know as a $T\overline{T}$ deformation. In 2D CFT a
 $T\overline{T}$ operator is an irrelevant operator where $T$ is the energy-momentum tensor. In holomorphic coordinates $z,\bar z$ the components are denoted by  $T = T_{zz}$ and $\overline{T} = T_ {\bar{z}\bar{z}}$ \cite{TT-2D,TT-2D-QFT}. Among the interesting properties of this irrelevant operator is that starting from an integrable QFT, the deformed theory preserve integrability \cite{zamolodchikov-integrable,canonical-integrability} and complete $1+1$ gravity  in the UV \cite{completation-UV}. An example is the $T\overline{T}$ deformation of the massless scalar field action that leads to Nambu-Goto action \cite{TT-2D-QFT}. Other examples and extensions can be found in  \cite{TT-SUSY-QM,TT-root,metric-approach,modmax-deform-2d,TT-QM}.  Another  interesting property of this deformation is that relates non/ultra-relativistic limits of the string sigma model \cite{Mapping-ultra/relat-TT,blair}. 
 
Denoting the deformed action as $S_{QFT} = S_{CFT} + S_{\gamma}$ where $S_\gamma = \gamma\int d^2x T\overline{T}$ and $\gamma$ is the deformation parameter,  the deformed action is a solution of the flow equation
  $\frac{dS_{QF T}}{d\gamma} = \int d^2x T\overline{T}$ for finite $\gamma$.
  
  By applying these technics, we can prove that ModMax is a deformation of Maxwell action \cite{Emergence-MM-TT,modmax-deform-2d}. 
  
  The role played by $\sqrt{T\overline{T}}$-deformations could have a relation with the realization of dualities in non linear theories \cite{TT-Sduality}.

		Identifying $\mathbf{E}\rightarrow\mathbf{\dot{q}},\mathbf{D}\rightarrow\mathbf{p},\mathbf{B}\rightarrow\mathbf{q}$,  we present here a $T\bar T$ deformation of a classical harmonic oscillator in 2D that shares many properties already present in ModMax theory. Duality symmetry is replaced by the transformation that rotates configuration variables with its conjugate momentum variables in the sense that $(\mathbf{p} \rightarrow -\mathbf{q},\mathbf{q} \rightarrow \mathbf{p})$. Lorentz symmetry is just represented by SO(2) space rotations and conformal symmetry restrict the form of the possible Hamiltonians (or Lagrangians) in a precise way that we will reveal later. The solution to these restrictions results in a dynamical system with a well defined Hamiltonian (or Lagrangian), which also is parametrized by $\gamma$ in the same way as in the ModMax theory.

	Either in phase space or configuration space, the equations of motion and couplings are quite complicated and seem to be very difficult to find an analytical solution to the system (see appendice \ref{Remarks-Lagrangian} and \ref{Remarks-Hamiltonian}). Nevertheless, a numerical study reveals that the solution space is quite symmetric, developing beautiful curves, that encourage us to try to find an analytical solution. Moreover, as we know that the $T\overline{T}$-deformations preserve integrability we worked out the analytical solution. 
	
		\begin{figure}[H]
			\centering
			\includegraphics[scale=.5]{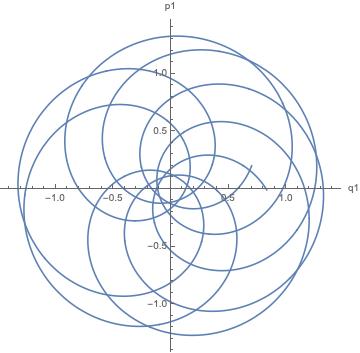}\hspace{1cm}
			\includegraphics[scale=.5]{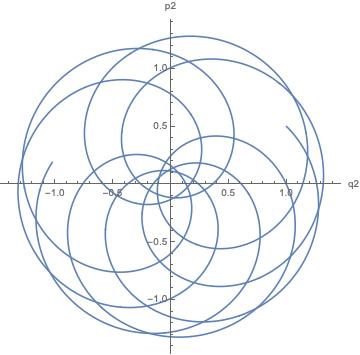}
			\caption{This solution is numerical and plotted using $\gamma=1$ and the initial conditions $q^1(0)=0.7$, $p^1(0)=0.2$, $q^2(0)=1.0$, $p^2(0)=0.5$, with time $t\in \{0,100\}$.}
			\label{solutions-plotted}
		\end{figure}
	
		 The Legendre transformation is also non-trivial as in the  ModMax theory.  In the classical mechanical system presented here the identification of two conserved quantities allows us to identify a relation between conjugate momenta and the configuration space variables $q, \dot{q}$ that is invertible resolving the explicit Legendre transformation. 
		
		In phase space, the generators of the infinitesimal duality rotations and the corresponding generator of space rotations are given respectively by 
		$$s=\frac{1}{2}(p^2+q^2),\hspace{1cm}j=\epsilon^{ij}p^iq^j,$$
		and are constants of motion. Using them we can parametrize the space of solutions in terms of two functions $A(s,j)$ and $B(s,j)$. Notice that we are not reducing the degrees of freedom but just parametrizing, in a different way, the space of solutions. This was a key observation that allowed us to find the analytical solution. The oscillator frequencies emerge under this parametrization as the functions $A$ and $B$. 
		
	
		Another interesting finding of this investigation is that  we can construct a map (depending on $q,p,\gamma$) of the 2D harmonic oscillator to the full non-linear oscillator at $2\gamma$.  
		
		A reformulation of the problem in configuration space starting from the ModMax Lagrangian presented in \cite{Kosyakov} can be constructed in terms of two constants of motion $C_1(\sigma, \rho)$, and $C_2(\sigma, \rho)$ where the variables $\sigma,\rho$ are defined by
		$$\sigma=\frac{1}{2}(\dot{q}^2+ q^2),\hspace{1cm}\rho=\epsilon^{ij}\dot{q}^iq ^j,$$
	
	The analytical results suggests an interesting physical interpretation. The couplings between oscillators corresponds to a oscillator mounted in a non-inertial oscillating frame. The frequencies of the oscillators are parametrized by the constants of motion $C_1$ and $C_2$. 
	
	We reveal also a phenomenon of beats related to the transfer of energy between the coupled oscillators and  we calculate the Hannay angle (that depends on the initial conditions through the numerical values of $C_1$ and $C_2$ evaluated on the given set of initial conditions). 
	
	All these interesting properties are rooted in the fact that we have a structure given by the duality symmetry implemented in the non-linear problem.
		
		The the paper is organized as follows. In the section \ref{TT-Lagrangian} we construct the non-linear model in the Lagrangian and Hamiltonian formalism. Later, in section \ref{Integration-Hamiltonian}, we integrate the Hamiltonian system, additionally we introduce the necessary notation to implement Legendre's transformation and integrate the Lagrangian system in the sections \ref{Legendre-Transform} and \ref{Integration-Lagrangian} respectively. In section \ref{Recursives-Relations}, we show how to construct the explicit map that transformation the 2D harmonic oscillator to the full non linear problem parametrized by $2\gamma$ and in the section \ref{Mechanical-Aspects}, we show that our model present a energy transfer phenomena between the oscillators and  compute the Hannay angle. Finally, in \ref{Conclusions}, we give our conclusions and the appendices  \ref{Remarks-Lagrangian} and \ref{Remarks-Hamiltonian} have complementary notes.	 
	
	\section{Non-linear oscillator from a $\sqrt{T\overline{T}}$-deformation} \label{TT-Lagrangian}
	
	Here we will construct the model using a $\sqrt{T\overline{T}}$-deformation. The model can be also constructed from scaling invariance (conformal invariance), see the appendices \ref{Remarks-Lagrangian} and \ref{Remarks-Hamiltonian}. It consists of taking the homogeneous harmonic oscillator in 2D and deforming it as ModMax. Even though this is the simpler model, we consider that its theoretical details are relevant to the understanding the non-linear systems that preserve duality and moreover improve the understanding of $T\overline{T}$-deformations. 
	
	\subsection{$\sqrt{T\overline{T}}$-deformations in Lagrangian formalism}

	We perform a $\sqrt{T\overline{T}}$-deformations for the harmonic oscillator. First, we consider the Lagrangian of the harmonic oscillator in $2D$ with masses $m_ i=1$ and frequencies $\omega_ i=1$ for $i=\{1,2\}$ 
	\begin{equation}
	L_0=\frac{1}{2}(\mathbf{\dot{q}}^2-\mathbf{q}^2),
	\end{equation} 
	where $\mathbf{\dot{q}}^2=\dot q_1^2+\dot q_2^2$ and $\mathbf{q}^2=q_1^2+q_2^2$.
	
	Due to the rotational and time translations invariance of the action, there exist two Noether conserved charges defined by
	\begin{equation}
	\label{E0}
	\delta_E q^i=\dot{q}^i, \hspace{1cm} E_0=\frac{1}{2}(\mathbf{\dot{q}}^2+\mathbf{q}^2),
	\end{equation}  
	\begin{equation}
	\label{J0}
	\delta_J q^i=\epsilon^{ij}q^j, \hspace{1cm} J_0=\epsilon^{ij}\dot{q}^iq^j.
	\end{equation}
	
	In order to preserve these symmetries at any order in $\gamma$, the parameter of the deformation, we define the conserved quantities $E_n$ and $J_n$ whose are the energy and angular momentum of the deformed action at order $n$. Then we define the $T\overline{T}$-like operator of order $n$
	\begin{equation}
	O^{\gamma}_{n}=\sqrt{E_n^2-J_n^2}.
	\end{equation}
	This is the analog to the $T\overline{T}$-like operator used in \cite{Emergence-MM-TT,modmax-deform-2d,TT-root} to deform the Maxwell into ModMax theory. 
	
	The deformed Lagrangian of order $n+1$ is defined by flow equation
	\begin{equation}
	L_{n+1}=L_0+\int d\gamma\  O_n^\gamma,
	\end{equation}
	then the first order deformation of the Lagrangian is
	\begin{equation}
	L_1=L_0+\gamma\sqrt{E_0^2-J_0^2},
	\end{equation}
	which is invariant under $\delta_E q^i$ and $\delta_J q^i$ symmetries. Then the deformed conserved quantities are
	\begin{equation}
	E_1=E_0\bigg(1+\frac{L_0\gamma}{\sqrt{E_0^2-J_0^2}}\bigg),
	\end{equation}
	\begin{equation}
	J_1=J_0\bigg(1+\frac{L_0\gamma}{\sqrt{E_0^2-J_0^2}}\bigg).
	\end{equation}
	
	Now, we are able to compute the operator $O^\gamma_1$ as
	\begin{equation}
	O^{\gamma}_{1}=\sqrt{E_1^2-J_1^2}=\sqrt{E_0^2-J_0^2}+\gamma L_0,
	\end{equation}
	thus the second order Lagrangian is
	\begin{equation}
	L_2=L_0+\gamma\sqrt{E_0^2-J_0^2}+\frac{1}{2}\gamma^2 L_0.
	\end{equation}
	
	Because of $\delta_E q^i$ and $\delta_J q^i$ are still symmetries of the action, we obtain the deformed second order conserved quantities
	\begin{equation}
	E_2=E_0\bigg(1+\frac{L_0\gamma}{\sqrt{E_0^2-J_0^2}}+\frac{1}{2}\gamma^2\bigg),
	\end{equation}
	\begin{equation}
	J_2=J_0\bigg(1+\frac{L_0\gamma}{\sqrt{E_0^2-J_0^2}}+\frac{1}{2}\gamma^2\bigg).
	\end{equation}
	and the second order $T\overline{T}$-like operator is
	\begin{equation}
	O_2^\gamma=\sqrt{E_0^2-J_0^2}+\gamma L_0,+\frac{1}{2}\gamma^2\sqrt{E_0^2-J_0^2},
	\end{equation}
	then the third order Lagrangian is
	\begin{equation}
	L_3=L_0+\gamma\sqrt{E_0^2-J_0^2}+\frac{1}{2}\gamma^2 L_0,+\frac{1}{3!}\gamma^3\sqrt{E_0^2-J_0^2}.
	\end{equation}
	
	By induction, it is straightforward to prove that the $n$ order deformation to the Lagrangian function is 
	\begin{equation}
	L=\sum_{n=0}^{\infty}\bigg[\bigg(\frac{\gamma^{2n}}{(2n)!} \bigg)L_0+\bigg(\frac{\gamma^{2n+1}}{(2n+1)!} \bigg)\sqrt{E_0^2-J_0^2}\bigg]=L_ 0\cosh\gamma+\sinh\gamma\sqrt{E_0^2-J_0^2},
	\end{equation}
	the $\delta_E q^i$ and $\delta_J q^i$ symmetries generate the conserved quantities \footnote{Compare this result with the Energy-momentum tensor of the ModMax theory
	$$T^{\mu\nu}=(F^{\mu\sigma}F^{\nu}{}_{\sigma}+ S\eta^{\mu\nu})\Big(\cosh\gamma+\sinh\gamma \frac{S}{\sqrt{S^2+P^2}}\Big).$$}
	\begin{equation}
	E=E_ 0\sum_{n=0}^{\infty}\bigg[\bigg(\frac{\gamma^{2n}}{(2n)!} \bigg)+\bigg(\frac{\gamma^{2n+1}}{(2n+1)!} \bigg)\frac{L_0}{\sqrt{E_0^2-J_0^2}} \bigg]=E_0\bigg(\cosh\gamma +\frac{L_0\sinh\gamma}{\sqrt{E_0^2-J_0^2}}\bigg),
	\label{deformed-energy-lagrangian}
	\end{equation}
	\begin{equation}
	J=J_ 0\sum_{n=0}^{\infty}\bigg[\bigg(\frac{\gamma^{2n}}{(2n)!} \bigg)+\bigg(\frac{\gamma^{2n+1}}{(2n+1)!} \bigg)\frac{L_0}{\sqrt{E_0^2-J_0^2}} \bigg]=J_0\bigg(\cosh\gamma +\frac{L_0\sinh\gamma}{\sqrt{E_0^2-J_0^2}}\bigg),
	\label{deformed-amomentum-lagrangian}
	\end{equation}
	it is important to stress that, the energy and the angular momentum of the theory are $E$ and $J$ respectively. The Hamiltonian function is NOT the energy of the dynamical system as we will see later.  
	
	To all orders $T\overline{T}$-like operator is
	\begin{equation}
	O^\gamma=\sum_{k=0}^{\infty}\bigg[\bigg(\frac{\gamma^{2n+1}}{(2n+1)!} \bigg)L_0+\bigg(\frac{\gamma^{2n}}{(2n)!} \bigg)\sqrt{E_0^2-J_0^2}\bigg]=L_ 0\sinh\gamma+\cosh\gamma\sqrt{E_0^2-J_0^2},
	\end{equation}
	which is solution of the $T\overline{T}$ flow equation
	\begin{equation}
	\frac{d S_{NL}}{d \gamma}=\int dt\ O^\gamma ,
	\end{equation}
	with $S_{NL}=\int dt\  L$.
	
	If we use the definitions
	\begin{equation}
	S=\frac{1}{2}(\mathbf{\dot{q}}^2-\mathbf{q}^2), \hspace{1cm} P=\mathbf{\dot{q}}\cdot\mathbf{q},
	\end{equation}
	we note that 
	\begin{equation}
	S=L_0,\hspace{1cm} E_0^2-J_0^2=S^2+P^2.
	\end{equation}
	Thus the deformed Lagrangian function is
	\begin{equation}
	L=S\cosh\gamma+\sinh\gamma\sqrt{S^2+P^2}.
	\label{Lagrangian-nlo}
	\end{equation}
	
	In terms of $E_0$ and $J_0$ we can write down the Lagrangian function as
	\begin{equation}
	L=E_ 0\cosh\gamma+\sinh\gamma\sqrt{E_0^2-J_0^2}-\mathbf{q}^2\cosh\gamma.
	\label{lagrangia-E-J}
	\end{equation}
	This form of the Lagrangian will be useful later when we implement the Legendre transformation to construct the Hamiltonian formalism.
	
	\subsection{$\sqrt{T\overline{T}}$-deformations in Hamiltonian formalism}
	
	Now we will  develop the $\sqrt{T\overline{T}}$-deformations in the Hamiltonian formalism.  To implement the deformation in Hamiltonian formalism we start with conserved quantities and then compute compute the Noether symmetries, so we use the deformed conserved quantities just constructed. Notice that the deformed energy and angular momentum are (\ref{deformed-energy-lagrangian}) and (\ref{deformed-amomentum-lagrangian}). In order to construct the deformations order by order in $\gamma$ as in the Lagrangian case, we start with the harmonic oscillator and its conserved quantities,
	\begin{equation}
	H_0=\frac{1}{2}(\mathbf{p}^2+\mathbf{q}^2),
	\end{equation}
	\begin{equation}
	s_0=\frac{1}{2}(\mathbf{p}^2+\mathbf{q}^2),\hspace{1cm}j_0=\epsilon^{ij}p^iq^j,
	\end{equation}
	where $\mathbf{p}^2=p_1^2+p_2^2$ and $\mathbf{q}^2=q_1^2+q_2^2$ and $i=\{1,2\}$. The conserved quantities $s_0$ and $j_0$ are just the Hamiltonian generators of the duality and rotational symmetries, 
	\begin{equation}
	\delta_{s_0}q^i=[q^i,s_0]=p^i,\hspace{1cm}\delta_{s_0}p^i=[p^i,s_0]=-q^i,
	\end{equation}
	\begin{equation}
	\delta_{j_0}q^i=[q^i,j_0]=\epsilon^{ij}q^j,\hspace{1cm}\delta_{j_0}p^i=[p^i,j_0]=-\epsilon^{ij}p^j.
	\end{equation}
	
	It is important to notice that $s_0$ coincide with the energy of the system at zero order in $\gamma$. Nevertheless at order $n>0$, $s_0$ is not the energy but is still the conserved quantity associated with the duality symmetry.  
	
	In order to keep $s_0$ and $j_0$ as Hamiltonian generators of the duality and rotational symmetry, we define the $T\overline{T}$-like Hamiltonian operator of order $n$ as
	\begin{equation}
	O^\gamma_n=\sqrt{s_n^2-j_n^2}.
	\end{equation}
The deformed Hamiltonian is defined as (flow equation)
	\begin{equation}
	H_{n+1}=s_0-\int d\gamma\ O^\gamma_n. 
	\end{equation}
	
	The first order deformation of the Hamiltonian function is then
	\begin{equation}
	H_1=s_0-\gamma \sqrt{s_0^2-j_0^2},
	\end{equation}
	 As a second step we propose  $s_1$ and $j_1$ in terms of $s_0$ and $j_0$ in the following way
	\begin{equation}
	s_1=s_0\bigg(1-\frac{s_0\gamma}{\sqrt{s_0^2-j_0^2}}\bigg),
	\end{equation}
	\begin{equation}
	j_1=j_0\bigg(1-\frac{s_0\gamma}{\sqrt{s_0^2-j_0^2}}\bigg).
	\end{equation}  
	$s_1$ and $j_1$ are conserved quantities of the system due to $s_0$ and $j_0$ are conserved. Using the Noether theorem, we compute the deformed symmetries  that are generated by $s_1$ and $j_1$ 
	 \begin{equation}
	 \delta_{s_1}q^i=\frac{\partial}{\partial p^i}\bigg(s_0\bigg[1-\frac{s_0\gamma}{\sqrt{s_0^2-j_0^2}}\bigg]\bigg),\hspace{1cm}	 \delta_{s_1}p^i=-\frac{\partial}{\partial q^i}\bigg(s_0\bigg[1-\frac{s_0\gamma}{\sqrt{s_0^2-j_0^2}}\bigg]\bigg),
	 \end{equation}  
	 \begin{equation}
	 \delta_{j_1}q^i=\frac{\partial}{\partial p^i}\bigg(j_0\bigg[1-\frac{s_0\gamma}{\sqrt{s_0^2-j_0^2}}\bigg]\bigg),\hspace{1cm}	 \delta_{j_1}p^i=-\frac{\partial}{\partial q^i}\bigg(j_0\bigg[1-\frac{s_0\gamma}{\sqrt{s_0^2-j_0^2}}\bigg]\bigg),
	 \end{equation} 
	 that correspond to duality and rotation deformed symmetries.
	
	Now we construct the first order $T\overline{T}$ operator
	\begin{equation}
	O_1^\gamma=\sqrt{s_0^2-j_0^2}-\gamma s_0,
	\end{equation}
	then the second order Hamiltonian function is
	\begin{equation}
	H_2=s_0-\gamma\sqrt{s_0^2-j_0^2}+\frac{1}{2}\gamma^2 s_0.
	\end{equation}
	
	In order to construct the second order $T\overline{T}$-like operator, we propose $s_2$ and $j_2$ conserved quantities in terms of $s_0$ and $j_0$ as folows
	\begin{equation}
	s_2=s_0\bigg(1-\frac{s_0\gamma}{\sqrt{s_0^2-j_0^2}}+\frac{1}{2}\gamma^2\bigg),
	\end{equation}
	\begin{equation}
	j_2=j_0\bigg(1-\frac{s_0\gamma}{\sqrt{s_0^2-j_0^2}}+\frac{1}{2}\gamma^2\bigg),
	\end{equation}  
	these conserved quantities generate the deformed symmetries
	\begin{equation}
	\delta_{s_1}q^i=\frac{\partial}{\partial p^i}\bigg(s_0\bigg[1-\frac{s_0\gamma}{\sqrt{s_0^2-j_0^2}}+\frac{1}{2}\gamma^2\bigg]\bigg),\hspace{1cm}	 \delta_{s_1}p^i=-\frac{\partial}{\partial q^i}\bigg(s_0\bigg[1-\frac{s_0\gamma}{\sqrt{s_0^2-j_0^2}}+\frac{1}{2}\gamma^2\bigg]\bigg),
	\end{equation}  
	\begin{equation}
	\delta_{j_1}q^i=\frac{\partial}{\partial p^i}\bigg(j_0\bigg[1-\frac{s_0\gamma}{\sqrt{s_0^2-j_0^2}}+\frac{1}{2}\gamma^2\bigg]\bigg),\hspace{1cm}	 \delta_{j_1}p^i=-\frac{\partial}{\partial q^i}\bigg(j_0\bigg[1-\frac{s_0\gamma}{\sqrt{s_0^2-j_0^2}}+\frac{1}{2}\gamma^2\bigg]\bigg).
	\end{equation}
	which in turn are the deformed duality and rotation symmetries.
	
	The second order $T\overline{T}$-like operator is
	\begin{equation}
	O_2^\gamma=\sqrt{s_0^2-j_0^2}-\gamma s_0+\frac{1}{2}\gamma^2 \sqrt{s_0^2-j_0^2},
	\end{equation}
	Continuing with the same steps, the third order Hamiltonian function is
	\begin{equation}
	H_3=s_0-\gamma\sqrt{s_0^2-j_0^2}+\frac{1}{2}\gamma^2 s_0-\frac{1}{3!}\gamma^3\sqrt{s_0^2-j_0^2}.
	\end{equation}
	
	By induction we prove that the Hamiltonian function at all orders in $\gamma$ is
	\begin{equation}
	H=\sum_{n=0}^{\infty}\bigg[\bigg(\frac{\gamma^{2n}}{(2n)!} \bigg)s_0-\bigg(\frac{\gamma^{2n+1}}{(2n+1)!} \bigg)\sqrt{s_0^2-j_0^2}\bigg]=s_ 0\cosh\gamma-\sinh\gamma\sqrt{s_0^2-j_0^2},
	\end{equation}
	which was constructed using the deformed energy and angular momentum $s$ and $j$ 
	\begin{equation}
	s=s_0\bigg(\cosh\gamma-\frac{s_0\sinh\gamma}{\sqrt{s_0^2-j_0^2}}\bigg),
	\end{equation}
	\begin{equation}
	j=j_0\bigg(\cosh\gamma-\frac{s_0\sinh\gamma}{\sqrt{s_0^2-j_0^2}}\bigg).
	\end{equation}  
	As a consequence of the underlaying scaling symmetry we found
	\begin{equation}
	\frac{s}{j}=\frac{s_0}{j_0}.
	\end{equation}
	These conserved quantities generate the deformed symmetries
		\begin{equation}
	\delta_{s}q^i=\frac{\partial}{\partial p^i}\bigg(s_0\bigg[\cosh\gamma-\frac{s_0\sinh\gamma}{\sqrt{s_0^2-j_0^2}}\bigg]\bigg),\hspace{1cm}	 \delta_{s}p^i=-\frac{\partial}{\partial q^i}\bigg(s_0\bigg[\cosh\gamma-\frac{s_0\sinh\gamma}{\sqrt{s_0^2-j_0^2}}\bigg]\bigg),
	\end{equation}  
	\begin{equation}
	\delta_{j}q^i=\frac{\partial}{\partial p^i}\bigg(j_0\bigg[\cosh\gamma-\frac{s_0\sinh\gamma}{\sqrt{s_0^2-j_0^2}}\bigg]\bigg),\hspace{1cm}	 \delta_{j}p^i=-\frac{\partial}{\partial q^i}\bigg(j_0\bigg[\cosh\gamma-\frac{s_0\sinh\gamma}{\sqrt{s_0^2-j_0^2}}\bigg]\bigg).
	\end{equation}

	The conserved quantities $s$ and $j$ can be used to construct  $T\overline{T}$-like operator that deforms the harmonic oscillators to the non-linear oscillators using the flow equation 
	$$H=-\int d\gamma\ O^\gamma \nonumber$$
	where
	\begin{equation}
		O^\gamma=-s_ 0\sinh\gamma+\cosh\gamma\sqrt{s_0^2-j_0^2},
	\end{equation}
	
	In the remaining of the article we will use instead of $s_0, j_0$ just $s,j$
	so the Hamiltonian function is
	\begin{equation}
	H=s\cosh\gamma-\sqrt{s^2-j^2}\sinh\gamma.
	\label{Hamiltonian-nlo}
	\end{equation}   
	Notice that this procedure generate the correct deformed Hamiltonian that can be constructed in same was as the ModMax Hamiltonian as presented in appendix \ref{Hamiltonian-ModMax}.   In contrast with the Lagrangian case here the symmetries and the constants of motion are deformed.
	
	\section{Integration of the Hamiltonian system} \label{Integration-Hamiltonian}
	
	Now we discuss how to use the conserved quantities $s$ and $j$ to integrate the Hamiltonian system. As is well known when we have $n$ degrees of freedom the Hamiltonian system needs $n$ conserved quantities in involution to be integrable \cite{Arnold}.  So from the knowledge of $s$ and $j$ we were able to integrate the system. 
	
	It is easy to prove that our non-linear oscillators  has two conserved quantities in involution that can be used to find an explicit analytic integration of the Hamiltonian equations of motion.  
	
	First notice that the Hamilton equations are
	 \begin{equation}
	\begin{split}
	\dot{p}^i&=-\frac{\partial H}{\partial q^i}=-\bigg(\cosh(\gamma)-\frac{\sinh(\gamma)s}{\sqrt{s^2-j^2}}\bigg) q^i+\frac{\sinh(\gamma)j}{\sqrt{s^2-j^2}}\epsilon^{ij}p^j,\\
	\dot{q}^i&=\frac{\partial H}{\partial p^i}=\bigg(\cosh(\gamma)-\frac{\sinh(\gamma)s}{\sqrt{s^2-j^2}}\bigg) p^i+\frac{\sinh(\gamma)j}{\sqrt{s^2-j^2}}\epsilon^{ij}q^j.
	\end{split}
	\label{hamiltonian-eqs}
	\end{equation}
	
	It is straightforward to prove that $s$ and $j$ are conserved quantities in involution, i.e
	\begin{equation}
	[s,j]=[s,H]=[j,H]=0.
	\end{equation}
	where $[F,G]$ denotes the Poisson bracket.
	
	Using this fact we can define two conserved quantities
	\begin{equation}
	A\equiv\cosh(\gamma)-\frac{\sinh(\gamma)s}{\sqrt{s^2-j^2}},\hspace{1cm}B\equiv\frac{\sinh(\gamma)j}{\sqrt{s^2-j^2}},
	\label{A-B-definitions-ps}
	\end{equation}
	that are also in involution
	\begin{equation}
	[A,B]=[A,H]=[B,H]=0.
	\end{equation}
	
	In terms of these quatities the Hamilton equations can be rewritten as
		 \begin{equation}
	\begin{split}
	\dot{p}^i&=-A q^i+B\epsilon^{ij}p^j,\\
	\dot{q}^i&=A p^i+B\epsilon^{ij}q^j,
	\end{split}
	\label{Hamiltonian-eq-A-B}
	\end{equation}
	Therefore, {\em over the surface defined by $A$ and $B$ constants}, the non-linear Hamiltonian equations are just linear equations! Moreover this reduction is consistent with the variational principle as we show in the appendices \ref{Remarks-Lagrangian} and \ref{Remarks-Hamiltonian}.
	
	We can rewritte the differential equations in a matrix notation as
	\begin{equation}
	\begin{pmatrix}
	\dot{p}^i \\
	\dot{q}^i
	\end{pmatrix}=\begin{pmatrix}
	B\epsilon^{ij} & -A\delta^{ij}\\
	A\delta^{ij} & B\epsilon^{ij}
	\end{pmatrix}\begin{pmatrix}
	p^j \\
	q^j
	\end{pmatrix},\hspace{1cm}C^{ab}\equiv\begin{pmatrix}
	B\epsilon^{ij} & -A\delta^{ij}\\
	A\delta^{ij} & B\epsilon^{ij}
	\end{pmatrix},\hspace{1cm}z^a\equiv\begin{pmatrix}
	p^j \\
	q^j
	\end{pmatrix},
	\end{equation}
	with $a,b=\{1,2,3,4\}$. Over the surface defined by $A$ and $B$ constants, the solutions of the system are obtained by taking the exponential of the matrix $C$  
	\begin{equation}
	\dot{z}^a=D^{ab}z_0^b,\hspace{1cm}D^{ab}=\big(e^{C t}\big)^{ab}
	\end{equation}
	where $z_0^b$ are the initial conditions.
	
	A remakable property of this dynamics is the fact the the matrix $C^{ab}$ admits a decomposition in term of two matrices $C_A$ and $C_B$ respectively
	\begin{equation}
	C^{ab}=\begin{pmatrix}
	B\epsilon^{ij} & -A\delta^{ij}\\
	A\delta^{ij} & B\epsilon^{ij}
	\end{pmatrix}=\begin{pmatrix}
	0 & -A\delta^{ij}\\
	A\delta^{ij} &0
	\end{pmatrix}+\begin{pmatrix}
	B\epsilon^{ij} & 0\\
	0 & B\epsilon^{ij}
	\end{pmatrix},
	\end{equation}
	These matrices commute between themselves and the exponential of $C $ is the exponential of $C_A $ times the exponential of $C_B $,  
	\begin{equation}
	D^{ab}=\big(e^{C_A t}e^{C_B t}\big)^{ab}=\big(e^{C_B t}e^{C_A t}\big)^{ab},\hspace{1cm} D_A\equiv e^{C_A t},\hspace{1cm}D_B\equiv e^{C_B t}.
	\end{equation}
	
	Then we observe that the dynamics of our system is the same as two coupled oscillators with frequencies $A$ and $B$ (that now we fix as numbers corresponding with the initial data). The explicit construction is then
	\begin{equation}
	D^{ab}_A z_0^b=
	\begin{pmatrix}
	p^i_0\cos (At)-q^i_0\sin (At) \\
	q^i_0\cos (At)+p^i_0\sin (At)
	\end{pmatrix},
	\end{equation}
	and
	\begin{equation}
	D^{ab}_B z_0^b=
	\begin{pmatrix}
	p^i_0\cos (Bt)+\epsilon^{ij}p^j_0\sin (Bt) \\
	q^i_0\cos (Bt)+\epsilon^{ij}q^j_0\sin (Bt)
	\end{pmatrix},
	\end{equation}
	Notice that the oscillations are not the same in phase-space.
	
	The solutions of the system are
	\begin{equation}
	\begin{split}
	p^i(t)=&\cos(At)(p_0^i\cos(Bt)+\epsilon^{ij}p_0^j\sin(Bt))-\sin(At)(q_0^i\cos(Bt)+\epsilon^{ij}q_0^j\sin(Bt)),\\
	q^i(t)=&\sin(At)(p_0^i\cos(Bt)+\epsilon^{ij}p_0^j\sin(Bt))+\cos(At)(q_0^i\cos(Bt)+\epsilon^{ij}q_0^j\sin(Bt)),
	\end{split}
	\label{Hamiltonian-solutions}
	\end{equation}
	where $p^i_0$ and $q^i_0$ are the initial conditions. These relations are the general solution of the system (\ref{hamiltonian-eqs}) and correspond to two coupled oscillators with frequencies $A$ and $B$ respectively.  
	
	\section{Legendre's transform} \label{Legendre-Transform}
	
	As has been shown in \cite{modmax-towsend}, Legendre's transformation in ModMax theory must be implemented carefully. Due to the non-linear behavior of the theory, the task of inverting the velocities in terms of momentum variables is cumbersome. A more easy task is to implement Legendre's transformation over the surface $A$ and $B$ constants because there the system is linear (cf. eq. (\ref{Hamiltonian-eq-A-B})). Moreover we know that a deduction of the corresponding Lagrangian  \cite{Kosyakov} can be implemented from first principles (appendix \ref{Lagrangian-ModMax}). 
	
	
	First, we consider the Hamiltonian function,
	\begin{equation}
	\label{HAB}
	H(q,p)=\frac{1}{2}A(q,p)(\mathbf{p}^2+\mathbf{q}^2)+B(q,p)\epsilon^{ij}p^iq^j,
	\end{equation}
	and we take $A$ and $B$ as constants. The Hamiltonian equations are 
	\begin{equation}
	\begin{split}
	\dot{p}^i&=-A q^i+B\epsilon^{ij}p^j,\\
	\dot{q}^i&=A p^i+B\epsilon^{ij}q^j,
	\end{split}
	\end{equation}
	by definition the substitution of $A$ and $B$ in terms of phase space variables in  (\ref{HAB}) leads to the non-linear Hamiltonian function (\ref{Hamiltonian-nlo}). A remarkable property of this Hamiltonian is that we can recover from it the complete equations of motion.  Now we can invert the momentum variables in terms of the configuration space
	\begin{equation}
	p^i=\frac{1}{A}\dot{q}^i-\frac{B}{A}\epsilon^{ij}q^j,
	\label{momentum-def-A-B-cts}
	\end{equation}	
	and over the surface defined by $A$ and $B$ constants,  compute the Lagrangian function
	\begin{equation}
	L=\frac{1}{2A}(\dot{\mathbf{q}}^2-(A^2-B^2)\mathbf{q}^2)-\frac{B}{A}\epsilon^{ij}\dot{q}^iq^j,
	\label{Lagrangian-A-B}
	\end{equation}
	of course, this Lagrangian functions is consistent with the momentum definition (\ref{momentum-def-A-B-cts}). 
	
	Now the question is how to write the corresponding expressions for $A$ and $B$ in terms of the configuration space. With the notation 
	\begin{equation}
	\sigma\equiv\frac{1}{2}(\mathbf{\dot{q}}^2+\mathbf{q}^2)=E_0,\hspace{1cm}\rho\equiv\epsilon^{ij}\dot{q}^iq^j=J_0,
	\label{sigma-rho}
	\end{equation}
	where $E_0$ and $J_0$ are in terms of $q,\dot q$ (eqs. (\ref{E0}),(\ref{J0})),
	we can define the two functions $C_1$ and $C_2$ 
	\begin{equation}
	C_1\equiv\cosh\gamma+\frac{\sigma\sinh\gamma}{\sqrt{\sigma^2-\rho^2}},\hspace{1cm}C_2\equiv-\frac{\rho\sinh\gamma}{\sqrt{\sigma^2-\rho^2}},
	\label{C1-C2}
	\end{equation}
	Now  the Lagrangian function (\ref{Lagrangian-nlo}) is 
	\begin{equation}
	L(\dot{\mathbf{q}},\mathbf{q})=C_1(\dot{\mathbf{q}},\mathbf{q})\sigma(\dot{\mathbf{q}},\mathbf{q})+C_2(\dot{\mathbf{q}},\mathbf{q})\rho(\dot{\mathbf{q}},\mathbf{q})-\mathbf{q}^2\cosh\gamma,
	\label{Lagrangian-C1-C2}
	\end{equation}
	Now we define the momentum by 
	\begin{equation}
	p^i=C_1(\dot{\mathbf{q}},\mathbf{q})\dot{q}^i+C_2(\dot{\mathbf{q}},\mathbf{q})\epsilon^{ij}q^j,
	\label{momentum-lagrangian-A-B-cts}
	\end{equation}
	This definition of the momentum is consistent with the starting Lagrangian in terms of $(q,\dot q)$. This property comes from the fact that the Lagrangian is a homogeneus function of degree one in $\sigma$ and $\rho$ (see appendix \ref{Remarks-Lagrangian} for details). 
	Demanding the consistency between the Hamiltonian momentum (\ref{momentum-def-A-B-cts}) and the Lagrangian momentum (\ref{momentum-lagrangian-A-B-cts}) we conclude that the implicit Legendre's transformation is
	\begin{equation}
	C_1(\dot{\mathbf{q}},\mathbf{q})=\frac{1}{A(\mathbf{p},\mathbf{q})},\hspace{1cm} C_2(\dot{\mathbf{q}},\mathbf{q})=-\frac{B(\mathbf{p},\mathbf{q})}{A(\mathbf{p},\mathbf{q})},
	\label{Legendre-transform-AB-C1C2}
	\end{equation}
	and using these identities, we can prove that the Lagrangian function (\ref{Lagrangian-A-B}) matches the Lagrangian function (\ref{Lagrangian-C1-C2}), the complete non-linear Lagrangian function.
	
	Therefore we have implemented Legendre's transformation over the surface defined by $A$ and $B$ constants. Due to $A$ and $B$ being constants of motion in phase space, $C_1$ and $C_2$ are also conserved in configuration space. This last statment can also be proved using the Lagrangian formalism. We want to stress that although $C_1$ and $C_2$ are conserved, $\sigma$ and $\rho$ are NOT conserved.
	
	\section{Integration of the Lagrangian system} \label{Integration-Lagrangian}
	
	Now we have all the necessary tools to show the complete  integration of the Lagrangian equations of motion. As in Hamiltonian formalism, we use the conserved quantities at the level of the equation of motion to integrate the system.
	
	The Lagrangian equations of motion are
	\begin{equation}
	\ddot{q}^i=\frac{C_2^2-1}{C_1^2}q^i-2\frac{C_2}{C_1}\epsilon^{ij}\dot{q}^j,
	\label{EM-Lagrangian}
	\end{equation}
	where we are using that $C_1$ and $C_2$ are constants of motion.
	
	The substitution of the expressions (\ref{C1-C2}) leads to the non-linear equation of motion of the Lagrangian function (\ref{Lagrangian-nlo}). More details about how the equations of motion can be reduced to this form  are given in the appendix \ref{Remarks-Lagrangian}. In particular, this differential equations are linear over the surface defined by $C_1$ and $C_2$ as constants, and the solution is
	\begin{equation}
	\begin{split}
	q^i(t)&=\cos\bigg(\frac{1}{C_1}t\bigg)\bigg[ q_0^i\cos\bigg(\frac{C2}{C1}t\bigg)-\epsilon^{ij}q_0^j\sin\bigg(\frac{C2}{C1}t\bigg) \bigg]\\&+\sin\bigg(\frac{1}{C_1}t\bigg)\bigg[\big(C_1\dot{q}_0^i+C_2\epsilon^{ij}q_0^j\big)\cos\bigg(\frac{C2}{C1}t\bigg)-\epsilon^{ij}\big(C_1\dot{q}^j_0+C_2\epsilon^{jk}q_0^k\big)\sin\bigg(\frac{C2}{C1}t\bigg)\bigg].
	\end{split}
	\label{Lagrangian-solutions}
	\end{equation}
	
	We notice that using the momentum's definition (\ref{momentum-def-A-B-cts}), the Hamiltonian solutions (\ref{Hamiltonian-solutions}) are consistent with (\ref{Lagrangian-solutions}). This fact is another check that our Legendre's transformation over the surface defined by $A$ and $B$ constants, or equivalently $C_1$ and $C_2$ as constants, is consistent.   
	
	Interestingly the functional form of the Hamiltonian and Lagrangian functions in terms of $(s,j)$ and $(S,P)$ respectively are the same as the Hamiltonian and Lagrangian densities of ModMax \cite{modmax-towsend, Kosyakov}. 
	
	
	\section{A deformation map} \label{Recursives-Relations}
	
	In this section we will show how to construct a map from the usual oscillator in 2D to the non-linear coupled oscillator presented here.  
	
	Starting from a 2D oscillator with mass and frequency equal to one
	\begin{equation}
	H_0=\frac12(\widehat p^2+\widehat q^2)
	\label{harmonic0}
	\end{equation}
	we define the new variables $\widehat p_i,
	\widehat q^i$ , using the matrix notation
		\begin{equation}
	\begin{pmatrix}
	\widehat p_i \\
	\widehat q^i
	\end{pmatrix}
	={\bold M}(\gamma)\begin{pmatrix}
	p^j \\
	 q^j
	\end{pmatrix},
	\label{Ham-matrix}
	\end{equation} 
	where
	\begin{equation}
	{\bold M}(\gamma)=
	\begin{pmatrix}
	B(q,p,\gamma)\epsilon^{ij}& -A(q,p,\gamma)\delta^{ij} \\
	 A(q,p,\gamma)\delta^{ij}& B(q,p,\gamma)\epsilon^{ij} 
	\end{pmatrix}.
	\end{equation}
	
	As $A$ and $B$ are functions of $q,p,\gamma$ the map is quite nonlinear. But on the surface $A$ and $B$ constants the map looks linear. This map is very powerful because, in just one step, implements the deformation of the harmonic oscillator (\ref{harmonic0}) into the full non-linear oscillator in $2\gamma$
	   \begin{equation}
	\label{HAB2gamma2}
	H(q,p,2\gamma)=\frac{1}{2}\Big(A^2(q,p,\gamma)+B^2(q,p,\gamma)\Big)(\mathbf{p}^2+\mathbf{q}^2)+2A(q,p,\gamma)B(q,p,\gamma)\epsilon^{ij}p^iq^j.
	\end{equation}
Doing the explicit calculation taking into account the functional dependence of $A$ and $B$ in terms of $,q,p,\gamma$ we find
	\begin{equation}
	\label{HAB2gamma}
	H(q,p,2\gamma)=\frac{1}{2}A(q,p,2\gamma)(\mathbf{p}^2+\mathbf{q}^2)+B(q,p,2\gamma)\epsilon^{ij}p^iq^j,
	\end{equation}
	      thus this map do the job of the $T\bar T$ deformation in just one step. But notice that here the deformation is obtained at $2\gamma$.
	
	
	Analogous  properties can be found for ModMax theory\footnote{ModMax satisfies a remarkable indentity that for the case of the oscillator considered here can be written in the form
	\begin{equation}
	H(\mathbf{p},\mathbf{q};2\gamma)=\frac{1}{2}\bigg[ \bigg(\frac{\partial H(\mathbf{p},\mathbf{q};\gamma)}{\partial p^i} \bigg)^2+ \bigg( \frac{H(\mathbf{p},\mathbf{q};\gamma)}{\partial q^i}\bigg)^2 \bigg].
	\label{identity-2gamma-Hamiltonian}
	\end{equation}
	The corresponding Lagranian identity is
	\begin{equation}
	L(\mathbf{\dot{q}},\mathbf{q};2\gamma)=-\frac{1}{2}\bigg[ \bigg(\frac{\partial L(\mathbf{\dot{q}},\mathbf{q};\gamma)}{\partial \dot{q}^i} \bigg)^2- \bigg( \frac{L(\mathbf{\dot{q}},\mathbf{q};\gamma)}{\partial q^i}\bigg)^2 \bigg].
	\end{equation}
	but here we will restrict ourselves to the Hamiltonian analysis.}. See \cite{modmax-towsend,Kosyakov} for details.
		
	
	In Lagrangian formalism the same idea can be implemented as follows.
	First we notice that the Lagrangian (\ref{Lagrangian-A-B}) in terms of $C_1$ and $C_2$ can be written as
	\begin{equation}
	L=\frac{1}{2}C_1\bigg(\mathbf{\dot{q}}^2+\frac{C_2^2-1}{C_1^2}\mathbf{q}^2 \bigg)+C_2\epsilon^{ij}\dot{q}^iq^j.
	\end{equation}
	 
	Using the matrix notation
		\begin{equation}
	L=\frac{1}{2}
	\begin{pmatrix}
	\dot{q}^i \\
	q^i
	\end{pmatrix}^T
	\begin{pmatrix}
	C_1\delta^{ij} & C_2\epsilon^{ij}\\
	 -C_2\epsilon^{ij} & \frac{C_2^2-1}{C_1}\delta^{ij}
	\end{pmatrix}\begin{pmatrix}
	\dot{q}^j \\
	 q^j
	\end{pmatrix}.
	\label{Lagrangian-matrix}
	\end{equation} 
	
	Now, we define the matrices and the vector
	\begin{equation}
	M^{ab}=\begin{pmatrix}
	C_1\delta^{ij} & C_2\epsilon^{ij}\\
	C_2\epsilon^{ij} & -\frac{C_2^2-1}{C_1}\delta^{ij}
	\end{pmatrix},\hspace{1cm}g^{ab}=\begin{pmatrix}
	\delta^{ij} & 0\\
	0& -\delta^{ij}
	\end{pmatrix},\hspace{1cm}v^a=\begin{pmatrix}
	\dot{q}^i \\
	q^i
	\end{pmatrix},
	\end{equation}
	$i={1,2}$ and $a,b={1,2,3,4}$. Using these definitions we have
	\begin{equation}
	L[v;2\gamma]=\frac{1}{2}(v^{T})^a(M^T(\gamma))^{ab}g^{bc}M^{cd}(\gamma)v^d,
	\label{Lagrangian-2gamma-M-v}
	\end{equation}
	where $M(\gamma)$ is the matrix $M$ evaluated on $\gamma$.
	Then we identify a new set of variables
	\begin{equation}
	\hat{v}^a=M(\gamma)^{ab}v^b,
	\label{variable-hat}
	\end{equation} 
	Using this new variables (\ref{variable-hat}) the Lagrangian (\ref{Lagrangian-2gamma-M-v}) is written as
	\begin{equation}
	L[\hat{v};2\gamma]=\frac{1}{2}(\hat{v}^T)^ag^{ab}\hat{v}^b,
	\end{equation}
	This last result is the Lagrangian of the harmonic oscillator in 2D. So we started with the full non linear Lagrangian and obtain through the definition (\ref{variable-hat}) the Lagrangian of the harmonic oscillator 2D.

	\section{Some dynamical properties of the system} \label{Mechanical-Aspects}
	In this section, we present two properties of the dynamical system. First, we show how these two coupled oscillators have an energy transfer phenomena between oscillators. Then we compute the Hannay angle, which is the classical analogue of the Berry phase for the corresponding quantum system. So the Hannay angle captures geometrical information (holonomy, parallel transport)  in the space of solutions.

	\subsection{Energy transfer }
	
	A well known property of  two coupled systems that oscillate and conserve energy, is the energy transfer phenomenon \cite{Landau}. It consists in that the amplitudes of the two oscillators also oscillate in such a way that when one of the systems oscillates with a large amplitude, the second one oscillates with a small amplitude and 
	viceversa.

	If we consider the amplitudes
	\begin{equation}
	\alpha^i= q_0^i\cos\bigg(\frac{C_2}{C_1}t\bigg)-\epsilon^{ij}q_0^j\sin\bigg(\frac{C_2}{C_1}t\bigg),
	\end{equation}
	and
	\begin{equation}
	\beta^i=\big(C_1\dot{q}_0^i+C_2\epsilon^{ij}q_0^j\big)\cos\bigg(\frac{C_2}{C_1}t\bigg)-\epsilon^{ij}\big(C_1\dot{q}^j_0+C_2\epsilon^{jk}q_0^k\big)\sin\bigg(\frac{C_2}{C_1}t\bigg),
	\end{equation}
 the solutions of the Lagrangian (\ref{Lagrangian-solutions}) system have the form
	\begin{equation}
	q(t)^i=\alpha^i \cos\bigg(\frac{1}{C_1}t\bigg)+\beta^i \sin\bigg(\frac{1}{C_1}t\bigg)
	\end{equation}
	which are oscillation functions that have amplitudes that also oscillate.
	
	 We can observe the transfer phenomenon if we plot some solutions of the system.
	
	\begin{center}
	\begin{figure}[H]
		\centering
		\includegraphics[scale=.5]{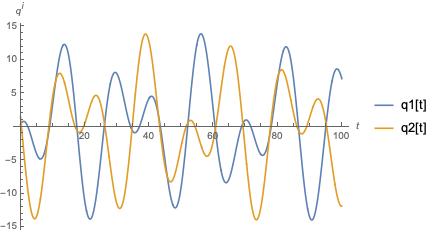}
		\caption{This solutions are plotted using $\gamma=1$ and the initial conditions $q^1(0)=1.5$, $\dot{q}^1(0)=-5.1$, $q^2(0)=0.3$, $\dot{q}^2(0)=1$.}
	\end{figure}
	\end{center}
	
	\subsection{Hannay angle}
	
	As we have observed, our system could be interpreted as an oscillator with frequency $A$ mounted in a non-inertial  reference frame that also oscillates but with frequency $B$. The Hannay angle \cite{Hannay} could be interpreted as a phase shift between the two oscillators.
	
	We compute the Hannay angle considering the shift $2\pi+Bt'$, where $t'$ is the time that takes one period of the first oscillator with frequency $A$. Then the Hannay angle in our system is
	\begin{equation}
	\Theta_H=2\pi+Bt'=2\pi\big(1+\frac{B}{A}\big), \hspace{1cm} t'=\frac{2\pi}{A},
	\end{equation}  
	
	Using the second equation in (\ref{Legendre-transform-AB-C1C2}), we can write the Hannay angle as
	\begin{equation}
	\Theta_H=2\pi\bigg(1-\frac{\rho\sinh\gamma}{\sqrt{\sigma^2-\rho^2}}\bigg).
	\end{equation}
	Of course, when $\gamma=0$, then $\Theta_H=2\pi$, it means that the two oscillators are in phase. This geometrical angle depends on the initial conditions through the definitions $A$ and $B$ evaluated at the corresponding initial condition vector. It also is the consequence of a coupling that can be interpreted as a covariant derivative
	\begin{equation}
	Dv^i=\frac{d v^i}{dt}+\frac{C_2}{C_1}\epsilon^{ij}q^j,
	\end{equation}
	then we can write the Lagrangian function (\ref{Lagrangian-A-B}) as
	\begin{equation}
	L=C_1Dq^i Dq^i-\frac{1}{C_1}q^2.
	\end{equation}

	\section{Conclusions} \label{Conclusions}
	
	We have constructed a non-linear classical system, which consists of two coupled oscillators. The construction can be performed by a $\sqrt{T\overline{T}}$-deformation (in Lagrangian and Hamiltonian formalisms) of the 2D homogeneous free harmonic oscillators. The system has a duality symmetry and conformal (scaling) symmetries. It could be interesting to study our system in the context of $T$-duality for point particles as \cite{klimcik}.
	
	The system is integrable, it has two conserved charges in involution $s$ and $j$, which are associated with duality and rotational symmetry respectively. It is very interesting that considering the constants of motion $A$ and $B$ (defined in terms of $s$ and $j$), it is straightforward to integrate the equation of motion as a linear system. Moreover, $A$ and $B$ are the frequencies of the two coupled oscillators respectively. Because we can use the conserved quantities in the action, we found how to perform the Legendre transformation in a simply way. It could be interesting to investigate the possibility of thinking of our model as a non-relativistic limit of some action, this research area has been addressed by \cite{Mapping-ultra/relat-TT,blair,JJ} in the context of $T\bar T$ deformations. Many of the dynamical properties of the system studied here can be translated to ModMax theory. This step is a work in progress that we will publish elsewhere.

	We construct a map of the 2D harmonic oscillator with mass and frequency equal to 1 to the nonlinear system at $2\gamma$. This map performs, in just one step, the complete deformation of the harmonic oscillator produced by the $T\bar T$ mechanism.

	We studied the mechanical properties of our system and found that the system present  energy transfer phenomenon. Finally, we compute the Hannay angle, which is a geometrical property of the system and space of solutions. 
	
	We leave for a future work the quantum analysis of the non linear  oscillator presented here.  Other applications and extensions like the supersymmetric case and the relativistic version are also worth to study in a future development of the ideas presented here. 
	
	\section*{Acknowledgements} 

RA was partially supported by a PhD. CONACyT fellowship number 744575. The work of JAG was partially supported by CONACyT grant A1-S-22886 and DGAPA-UNAM grant IN107520.

	\begin{appendices}
	\section{Remarks on the Lagrangian formalism} \label{Remarks-Lagrangian}
	
	In this appendix, we construct, from first principles our dynamical system showing the role played by the conserved quantities  $C_1$ and $C_2$. 
	
	Starting from the Lagrangian  
	\begin{equation}
	L=\sigma\cosh\gamma+\sqrt{\sigma^2-\rho^2}\sinh\gamma-\mathbf{q}^2\cosh\gamma,
	\end{equation} 
	using $\sigma$ and $\rho$ as in (\ref{sigma-rho}), we define
	\begin{equation}
	\bar{L}=\sigma\cosh\gamma+\sqrt{\sigma^2-\rho^2}\sinh\gamma,
	\end{equation}
	and observe that $\bar{L}$ is an homogeneous function of degree $1$ in $(\sigma,\rho)$, son we have the relation\footnote{Notice that this relation is the expression of scaling (conformal) symmetry as in the ModMax theory.}
	\begin{equation}
	\bar{L}(\sigma,\rho)=\frac{\partial \bar{L}}{\partial \sigma}\sigma+\frac{\partial \bar{L}}{\partial \rho}\rho.
	\end{equation}
	The complete Lagrangian function is
	\begin{equation}
	L=\frac{\partial \bar{L}}{\partial \sigma}\sigma+\frac{\partial \bar{L}}{\partial \rho}\rho-\mathbf{q}^2\cosh\gamma.
	\end{equation}
	
	From the definition of conjugate  momenta 
	\begin{equation}
	p^i=\frac{\partial^2 L}{\partial \dot{q}^i\partial \sigma}\sigma+\frac{\partial^2 L}{\partial \dot{q}^i\partial \rho}\rho+\frac{\partial L}{\partial \sigma}\frac{\partial\sigma}{\partial \dot{q}^i}+\frac{\partial L}{\partial \rho}\frac{\partial\rho}{\partial \dot{q}^i},
	\end{equation}
	we can  prove that
	\begin{equation}
	\frac{\partial^2 L}{\partial \dot{q}^i\partial \sigma}\sigma+\frac{\partial^2 L}{\partial \dot{q}^i\partial \rho}\rho=0.
	\end{equation}
	
	Therefore, if we define
	\begin{equation}
	C_1=\frac{\partial \bar{L}}{\partial \sigma},\hspace{1cm}C_2=\frac{\partial \bar{L}}{\partial \rho},
	\end{equation}
	we obtain the expressions for the Lagrangian and the momenta respectively
	\begin{equation}
	L=C_1(\mathbf{\dot{q}},\mathbf{q})\sigma(\mathbf{\dot{q}},\mathbf{q})+C_2(\mathbf{\dot{q}},\mathbf{q})\rho(\mathbf{\dot{q}},\mathbf{q})-\mathbf{q}^2\cosh\gamma,
	\end{equation} 
	\begin{equation}
	p^i=C_1(\mathbf{\dot{q}},\mathbf{q})\dot{q}^i+C_2(\mathbf{\dot{q}},\mathbf{q})\epsilon^{ij}q^j
	\end{equation}
	where it seems like if we can derive respect to $\dot{q}^i$ without taking into account the dependence of $C_1$ and $C_2$ on $\dot q^i$.
	
	The equations of motions are
	\begin{equation}
	\dot{C_1}\dot{q}^i+\dot{C_2}\epsilon^{ij}q^j+C_1\ddot{q}^i+C_2\epsilon^{ij}\dot{q}^j=\frac{\partial C_1}{\partial q^i}\sigma+\frac{\partial C_2}{\partial q^i}\rho+\frac{C_2^2-1}{C_1}q^i-C_2\epsilon^{ij}\dot{q}^j,
	\end{equation} 	
	Taking in to account  the time derivative of $C_1$ and $C_2$, we can collect the terms with $\ddot{q}^i$ 
	\begin{equation}
	\bigg(C_1\delta^{ik}+\frac{\partial C_1}{\partial \dot{q}^k}\dot{q}^i+\frac{\partial C_2}{\partial \dot{q}^k}\epsilon^{ij}q^j \bigg)\ddot{q}^k=-\frac{\partial C_1}{\partial q^k}\dot{q}^k\dot{q}^i-\frac{\partial C_2}{\partial q^k}\dot{q}^k\epsilon^{ij}q^j+\frac{\partial C_1}{\partial q^i}\sigma+\frac{\partial C_2}{\partial q^i}\rho+\frac{C_2^2-1}{C_1}q^i-2C_2\epsilon^{ij}\dot{q}^j,
	\end{equation}
	and identify
	\begin{equation}
	W^{ik}=C_1\delta^{ik}+\frac{\partial C_1}{\partial \dot{q}^k}\dot{q}^i+\frac{\partial C_2}{\partial \dot{q}^k}\epsilon^{ij}q^j.
	\end{equation}
	
	It is straightforward to prove that 
	\begin{equation}
	\frac{\partial C_1}{\partial q^i}\sigma+\frac{\partial C_2}{\partial q^i}\rho=0,
	\end{equation}
	then the equations of motion are
	\begin{equation}
	W^{ik}\ddot{q}^k=-\frac{\partial C_1}{\partial q^k}\dot{q}^k\dot{q}^i-\frac{\partial C_2}{\partial q^k}\dot{q}^k\epsilon^{ij}q^j+\frac{C_2^2-1}{C_1}q^i-2C_2\epsilon^{ij}\dot{q}^j,
	\end{equation}
	This equations can be simplified drastically over the surface over the surface $C_1, C_2$ constants  
	\begin{equation}
	\ddot{q}^i=\frac{C_2^2-1}{C_1^2}q^i-2\frac{C_2}{C_1}\epsilon^{ij}\dot{q}^j.
	\end{equation}

 	\section{Remarks on the Hamiltonian formalism} \label{Remarks-Hamiltonian}
 	
 	In the Hamiltonian formalism some aspects can be simplified. The conformal condition, as in ModMax, implies that the Hamiltonian function (\ref{Hamiltonian-nlo}) is a homogeneous function of degree $1$ in terms of $(s,j)$, 
 	\begin{equation}
 	H(s,j)=\frac{\partial H}{\partial s}s+\frac{\partial H}{\partial j}j.
 	\end{equation}

 	Then the Hamiltonian equations are
 	\begin{equation}
 	\dot{q}^i=\frac{\partial^2 H}{\partial p^i\partial s}s+\frac{\partial^2 H}{\partial p^i\partial j}j+\frac{\partial H}{\partial s}\frac{\partial s}{\partial p^i}+\frac{\partial H}{\partial j}\frac{\partial j}{\partial p^i},
 	\end{equation}
 	\begin{equation}
 	\dot{p}_i=-\bigg[\frac{\partial^2 H}{\partial q^i\partial s}s+\frac{\partial^2 H}{\partial q^i\partial j}j+\frac{\partial H}{\partial s}\frac{\partial s}{\partial q^i}+\frac{\partial H}{\partial j}\frac{\partial j}{\partial q^i}\bigg].
 	\end{equation}
 	
 	It is straightforward to show that, the first two terms in each Hamiltonian equations satisfy
 	\begin{equation}
 	\frac{\partial^2 H}{\partial p^i\partial s}s+\frac{\partial^2 H}{\partial p^i\partial j}j=0,\hspace{1cm} \frac{\partial^2 H}{\partial q^i\partial s}s+\frac{\partial^2 H}{\partial q^i\partial j}j=0,
 	\end{equation}
 	therefore, if we define  
 	\begin{equation}
 	A=\frac{\partial H(s,j)}{\partial s}\hspace{1cm}B=\frac{\partial H(s,j)}{\partial j},
 	\end{equation} 	
 	then we obtain 
 	\begin{equation}
 	H(\mathbf{q},\mathbf{p})=A(\mathbf{q},\mathbf{p})s(\mathbf{q},\mathbf{p})+B(\mathbf{q},\mathbf{p})j(\mathbf{q},\mathbf{p}),
 	\end{equation}
 	and the Hamiltonian equations are as in (\ref{Hamiltonian-eq-A-B}). 
 	
 	Of course, as we have shown in the main text, $A$ and $B$ are conserved quantities of the system, but the interesting remark is that we can derive the Hamiltonian equations by explicity take into account the derivatives of $A$ and $B$ as functions of $(\mathbf{q},\mathbf{p})$, or we can obtain the Hamiltonian equations just by taking $A$ and $B$ as constants. The two procedures gives the same result as in the equations (\ref{A-B-definitions-ps}).
 	
 	\section{Hamiltonian deduction of ModMax}
 	\label{Hamiltonian-ModMax}
 	
 	We will summarize some relevant results from the Hamiltonian deduction of ModMax. The Hamiltonian density $\mathcal{H}$ for a generic sourceless electromagnetic theory in the vacuum must be dependent on magnetic induction 3-vector $\textbf{B}$ and the displacement current 3-vector $\textbf{D}$. The equations of motion are
 	\begin{equation}
 	\begin{split}
 	\dot{\textbf{B}}&=-\nabla\times \textbf{E}, \hspace{1cm} \nabla\cdot \textbf{B}=0,
 	\\
 	\dot{\textbf{D}}&=\nabla\times \textbf{H}, \hspace{1.25cm} \nabla\cdot \textbf{D}=0,
 	\end{split}
 	\label{maxwell-equations}
 	\end{equation}
 	taken together with the constitutive relations
 	\begin{equation}
 	\textbf{E}=\frac{\partial \mathcal{H}}{\partial \textbf{D}}, \hspace{1cm} \textbf{H}=\frac{\partial \mathcal{H}}{\partial \textbf{B}}.
 	\label{constitutive-relations}
 	\end{equation}
 	
 	The equations of motion are invariant under time and spatial translations implying that, the integrals over space of
 	\begin{equation}
 	\dot{\mathcal{H}}=-\nabla\cdot(\textbf{E}\times\textbf{H}), \hspace{1cm}\mathcal{P}_i=-\partial_jT^j_{\ i}, 
 	\label{H-punto-P-punto}
 	\end{equation}
 	are conserved charges, where $\mathcal{P}_i$, $i=\{1,2,3\}$, are the components of the density momentum $\mathbfcal{P}=\textbf{D}\times\textbf{B}$ and $T^j_{\ i}$ are the components of $3\times3$ energy-momentum tensor,
 	\begin{equation}
 	T^i_{\ j}=\delta^i_{\ j}(\textbf{D}\cdot\textbf{E}+\textbf{H}\cdot\textbf{B}-\mathcal{H})-(D^iE_j+H^iB_j),
 	\end{equation}
 	taking into account the rotational invariance, these are the manifest symmetries of the field equations. The rotational invariance implies $\mathbf{B}\times \mathbf{H}+\mathbf{D}\times\mathbf{E}=0$.
 	
 	There are other symmetries that are not manifest in the field equations, such as Lorentz boost invariance. In a Lorentz invariant theory is possible to write the equations (\ref{H-punto-P-punto}) as the 4-vector continuity equation for a symmetric energy momentum tensor, if only if,
 	\begin{equation}
 	\textbf{E}\times\textbf{H}=\textbf{D}\times\textbf{B},
 	\label{lorentz-invariant-condition}
 	\end{equation}
 	which is therefore the condition for the equations (\ref{maxwell-equations}) to be Lorentz invariant. 
 	
 	The trace of the energy-momentum tensor is
 	\begin{equation}
 	T^i_{\ i}-\mathcal{H}=2[\textbf{D}\cdot\textbf{E}+\textbf{H}\cdot\textbf{B}-2\mathcal{H}],
 	\end{equation}
 	thus the conditions of conformal invariance are (\ref{lorentz-invariant-condition}) and
 	\begin{equation}
 	\textbf{D}\cdot\textbf{E}+\textbf{H}\cdot\textbf{B}=2\mathcal{H},
 	\label{conformal-condition}
 	\end{equation}
 	this last equation, in terms of the constitutive equations (\ref{constitutive-relations}), means that the Hamiltonian density is a homogeneous function of degree $2$ in terms of $(\mathbf{D},\mathbf{B})$. 
 	
 	Finally, the condition for invariance under the $SO(2)$ electromagnetic duality, which acts as rotations between the 3-vectors $\textbf{D}$ and $\textbf{B}$, is
 	\begin{equation}
 	\textbf{E}\cdot\textbf{B}=\textbf{D}\cdot\textbf{H}.
 	\label{hamiltonian-condition-duality}
 	\end{equation}
 	Although, the are 3 independent rotation scalars, there are at most two duality invariants
 	\begin{equation}
 	s=\frac{1}{2}(\textbf{D}^2+\textbf{B}^2),\hspace{1cm}p=|\textbf{D}\times\textbf{B}|.
 	\end{equation}
 	If $\mathcal{H}$ is a duality invariant, it must be a function of $s$ and $p$. The Lorentz invariant condition (\ref{lorentz-invariant-condition}) implies, using (\ref{constitutive-relations}), that
 	\begin{equation}
 	\mathcal{H}_s^2+\frac{2s}{p}\mathcal{H}_s\mathcal{H}_p+\mathcal{H}_p^2=1.
 	\label{ecuacion-para-Hamiltoniano}
 	\end{equation}
 	
 	An alternative basis for the duality invariant rotation scalars is
 	\begin{equation}
 	u=\frac{1}{2}\bigg(s+\sqrt{s^2-p^2} \bigg), \hspace{1cm}v=\frac{1}{2}\bigg(s-\sqrt{s^2-p^2} \bigg).
 	\end{equation}
 	These variables are well-defined since,
 	\begin{equation}
 	s^2-p^2=\xi^2+\eta^2\geq 0,
 	\end{equation}
 	where $(\xi,\eta)$ are rotation invariants,
 	\begin{equation}
 	\xi=\frac{1}{2}(\mathbf{D}^2-\mathbf{B}^2),\hspace{1cm}\eta=\mathbf{D}\cdot\mathbf{B},
 	\end{equation}
 	notice that $(\xi,\eta)$ are not duality invariants.
 	
 	For solution of the form $\mathcal{H}=\sqrt{K}+\text{constant}$, the equation (\ref{ecuacion-para-Hamiltoniano}) in terms of $(u,v)$ variables is $K_uK_v=4K$. The solution for quadratic and non-negative $K(u,v)$ and zero vacuum energy depend on one parameter $T$ with dimensions of energy density and a dimensionless parameter $\gamma$,
 	\begin{equation}
 	\mathcal{H}=\sqrt{T^2+2T(e^{-\gamma}u+e^{\gamma}v)+4uv}-T.
 	\end{equation}
	The Lagrangian version was constructed in \cite{p-form}.
 	This  Hamiltonian density is the Born-Infeld electrodynamics \cite{BI}, when $\gamma=0$. The strong field limit, $T\rightarrow 0$, yields the duality and conformal invariant Hamiltonian density $\mathcal{H}=p$ of Bialynicki-Birula electrodynamics \cite{BB-1,BB-2}, for all $\gamma$. The attempt to find a Lagrangian density fails, since $\mathbf{D}\cdot\mathbf{E}-\mathcal{H}\equiv 0$. The week field limit, $T\rightarrow \infty$ yields the Hamiltonian density
 	\begin{equation}
 	\mathcal{H}=(\text{cosh}\gamma) s-(\text{sinh}\gamma)\sqrt{s^2-p^2},
 	\label{ModMax-s,p}
 	\end{equation}  
 	an equivalent expression is,
 	\begin{equation}
 	\mathcal{H}=\frac{1}{2}(\text{cosh}\gamma)(\mathbf{D}^2+\mathbf{B}^2)-\frac{1}{2}(\text{sinh}\gamma)\sqrt{(\mathbf{D}^2-\mathbf{B}^2)^2+4(\mathbf{D}\cdot\mathbf{B})^2},
 	\label{ModMax}
 	\end{equation}
 	and
 	\begin{equation}
 	\mathcal{H}=\frac{1}{2}(\text{cosh}\gamma)(\mathbf{D}^2+\mathbf{B}^2)-\frac{1}{2}(\text{sinh}\gamma)\sqrt{(\mathbf{D}^2+\mathbf{B}^2)^2-4(\mathbf{D}\times\mathbf{B})^2}.
 	\end{equation}
 	This is the one-parameter extension of Maxwell electrodynamics. For any value of $\gamma$, this Hamiltonian density satisfies (\ref{conformal-condition}) required for conformal invariance. Although, there may be other non-linear Lorentz and duality invariant theories of electrodynamics corresponding to other solutions of (\ref{ecuacion-para-Hamiltoniano}), there exist just two theories that are also conformal invariant, Bialynicki-Birula electrodynamics and the family of ModMax theories as \cite{modmax-towsend} has proven. 
 	
 	\section{Lagrangian deduction of ModMax}
 	\label{Lagrangian-ModMax}
 	
 	The existence of the ModMax Lagrangian density can be derived in a simple way \cite{Kosyakov}.  
 	
 	Considering the Bessel-Hagen criterion for conformal invariance is \cite{bessel-hagel} 
 	\begin{equation}
 	\Theta^{\mu}_{\ \mu}=0,
 	\label{beesel-hagen}
 	\end{equation}
 	where $\Theta^{\mu}_{\ \mu}$ is the symmetric $4\times4$ energy-momentum tensor.
 	
 	The Lagrangian density must be a functions of $(S,P)$, in this case the equation (\ref{beesel-hagen}) can be cast as
 	\begin{equation}
 	\mathcal{L}_SS+\mathcal{L}_PP=\mathcal{L},
 	\label{lagrangiano-conforme}
 	\end{equation}
 	where $S=-\frac{1}{4}F_{\mu\nu}F^{\mu\nu}$, $P=-\frac{1}{4}F_{\mu\nu}\tilde{F}^{\mu\nu}$ are the Lorentz invariants, together with the definitions $\mathcal{L}_SS=\partial \mathcal{L}/\partial S$ and $\mathcal{L}_PP=\partial \mathcal{L}/\partial P$. Again, this last equation means that the Lagrangian density is a homogeneous function of degree $1$ in terms of $(S,P)$. 
 	
 	The equations of motion are
 	\begin{equation}
 	\partial_\mu E^{\mu\nu}=0 ,
 	\end{equation}
 	where 
 	\begin{equation}
 	E_{\mu\nu}=\frac{\partial\mathcal{L}}{\partial F^{\mu\nu}}=-(\mathcal{L}_SF_{\mu\nu}+\mathcal{L}_P  \tilde{F}_{\mu\nu} ),
 	\label{E-del-ruso}
 	\end{equation}
 	with the Bianchi identity
 	\begin{equation}
 	\partial_\mu \tilde{F}^{\mu\nu}=0,
 	\end{equation}
 	which can be used to solve $F_{\mu\nu}$ in terms of the 1-form gauge potential $A_{\mu}$.
 	
 	In this Lagrangian approach, the constitutive equations are defined through (\ref{E-del-ruso}). In general, they are not duality invariant. The Gaillard-Zumino criterion \cite{galliard-zumino} for invariance under duality-rotation transformations is
 	\begin{equation}
 	\tilde{E}_{\mu\nu}E^{\mu\nu}=\tilde{F}_{\mu\nu}F^{\mu\nu},
 	\label{duality-condition-lagrangian}
 	\end{equation}
 	which is equivalent to equation (\ref{hamiltonian-condition-duality}).
 	
 	Using (\ref{E-del-ruso}) and the identity $\tilde{F}_{\mu\nu}\tilde{F}^{\mu\nu}=-F_{\mu\nu}F^{\mu\nu}$, the equation  (\ref{duality-condition-lagrangian}) can be rewrite as follows
 	\begin{equation}
 	4\bigg(\mathcal{L}_S^2-\mathcal{L}_P^2\bigg)P-8\mathcal{L}_S\mathcal{L}_PS=P.
 	\label{ecuacion-partial diferential-lagrangiana}
 	\end{equation} 
 	multiplying both sides of the equation (\ref{ecuacion-partial diferential-lagrangiana}) and using (\ref{lagrangiano-conforme}), it result in 
 	\begin{equation}
 	4\bigg(\sqrt{S^2+P^2}\mathcal{L}_S-\mathcal{L}\bigg)\bigg(\sqrt{S^2+P^2}\mathcal{L}_S+\mathcal{L} \bigg)=P^2.
 	\label{ecuacion diferencial partial lagrangian 2}
 	\end{equation}
 	
 	To solve this non-linear partial differential equation, it is convenient to define,
 	\begin{equation}
 	x=\sqrt{S^2+P^2},\hspace{1cm}y=S,
 	\end{equation}
 	this  $x$ and $y$ are independent variables everywhere except for the point $P=0$ which is the only singular point of the Gaillard-Zumino condition (\ref{duality-condition-lagrangian}).
 	
 	Since  equation (\ref{lagrangiano-conforme}) implies  that $\mathcal{L}$ is homogeneus  of degree 1 in $S$ and $P$, Searching for solutions of  (\ref{ecuacion diferencial partial lagrangian 2}) in the form,
 	\begin{equation}
 	\mathcal{L}=\alpha x+\beta y,
 	\label{ansatz}
 	\end{equation}
 	with $\alpha$ and $\beta$ to be determined and substituting this proposal in equation (\ref{ecuacion diferencial partial lagrangian 2}), the values for $x$ and $y$ can be obtained,
 	\begin{equation}
 	\alpha=\pm \sinh \gamma,\hspace{1cm}\beta=\cosh\gamma.
 	\end{equation}
 	
 	The solution with $\alpha=-\sinh$, $\beta=\cosh\gamma$, $\gamma>0$ represent the Lagrangian density which is unbounded from below and should be discarded. Hence the set of Lagrangian densities, invariant under conformal group transformation and duality rotations, is given by the one-parameter family of functions,
 	\begin{equation}
 	\mathcal{L}(S,P;\gamma)=(\cosh\gamma)S+(\sinh\gamma)\sqrt{S^2+P^2},
 	\end{equation} 
 	where $\gamma\geq 0$.
 	
	\end{appendices}

\end{document}